\documentstyle[eqsecnum,cite,floats,epsfig,amssymb,aps,prb,preprint]{revtex}

\newcommand{\psfigure}[1]{\centerline{\psfig{file=#1,width=13cm}}}

\tightenlines 
\begin{document}
\author{Dirk Reith\footnote{corresponding author, email: 
\verb+reith@mpip-mainz.mpg.de+}, 
Hendrik Meyer and Florian M\"uller-Plathe}
\address{
  Max-Planck-Institut f\"ur Polymerforschung,
  Ackermannweg 10, D-55128 Mainz, Germany}
\title{Mapping atomistic to coarse-grained polymer models
using automatic simplex optimization to fit structural properties}
\date{\today}
\maketitle
%
%
%
\begin{abstract}
We develop coarse-grained force fields for poly (vinyl alcohol) and poly 
(acrylic acid) oligomers. In both cases, one monomer is mapped onto 
a coarse-grained bead. The new force fields are designed to match structural 
properties such as radial distribution functions of various kinds
derived by atomistic simulations of these polymers. 
The mapping is therefore constructed in a way to take into account as much
atomistic information as possible.
On the technical side, our approach consists of a simplex algorithm which is
used to optimize automatically non-bonded parameters as well as bonded
parameters. Besides their similar conformation
(only the functional side group differs), poly (acrylic acid) was chosen to
be in aqueous solution in contrast to a poly (vinyl alcohol) melt. For poly 
(vinyl alcohol) a non-optimized bond angle potential turns out to be 
sufficient in connection with a special, optimized non-bonded potential. 
No torsional potential
has to be applied here. For poly (acrylic acid), we show that each peak of 
the radial distribution function is usually dominated by some specific model
parameter(s). Optimization of the bond angle parameters is essential. 
The coarse-grained forcefield reproduces the radius of gyration
$R_{G}$ of the atomistic model. As a first application, we use the force 
field to simulate longer chains and compare the hydrodynamic
radius $R_{H}$ with experimental data.
\end{abstract}

\section{Introduction}
Many computer simulations of polymers suffer from the detailed treatment of 
the fast modes (e.g. bond vibrations) on the microscopic length scale: It has
the serious consequence that the time propagation of the system can only be
performed in tiny steps. Even the application of supercomputers can not
overcome the essential problem that the total simulated time of the system is
typically limited to the order of nanoseconds. Consequently, the slow modes
(e.g. radius of gyration or end-to-end vector) can hardly be equilibrated. 
It is therefore not
feasible to perform fully detailed atomistic simulations in order to derive
specific macroscopic bulk properties of polymers. The study of realistically 
large atomistic systems is far from being feasible.\cite{baschnagel00}

This study is geared to the question of how to enhance size and simulated
real time of a polymer system in computer simulations. Observe that we refer
to specific polymer systems with detailed information on the atomistic length 
scale, not just to generic Gaussian chains. Such an enhancement would 
especially be useful if the resulting simulations would still keep an essential
part of the atomistic information of the system. 

In recent years, various so-called coarse-graining (CG) methods have been 
developed 
to enhance polymer simulations. They are very different in spirit and purpose.
Some examples are the `time coarse graining',\cite{forrest95} lattice 
approaches like the bond fluctuation model,
\cite{carmesin88,paul91,baschnagel91} or the high coordination 
lattice.\cite{doruker97,doruker99a} On a larger length scale,
dissipative particle dynamics (DPD) and smoothed particle dynamics (SPD)
are frequently used to tackle hydrodynamic problems.\cite{groot97,espanol97}

\begin{figure}[hb]
  \begin{center}
    \epsfxsize 15cm \epsfbox {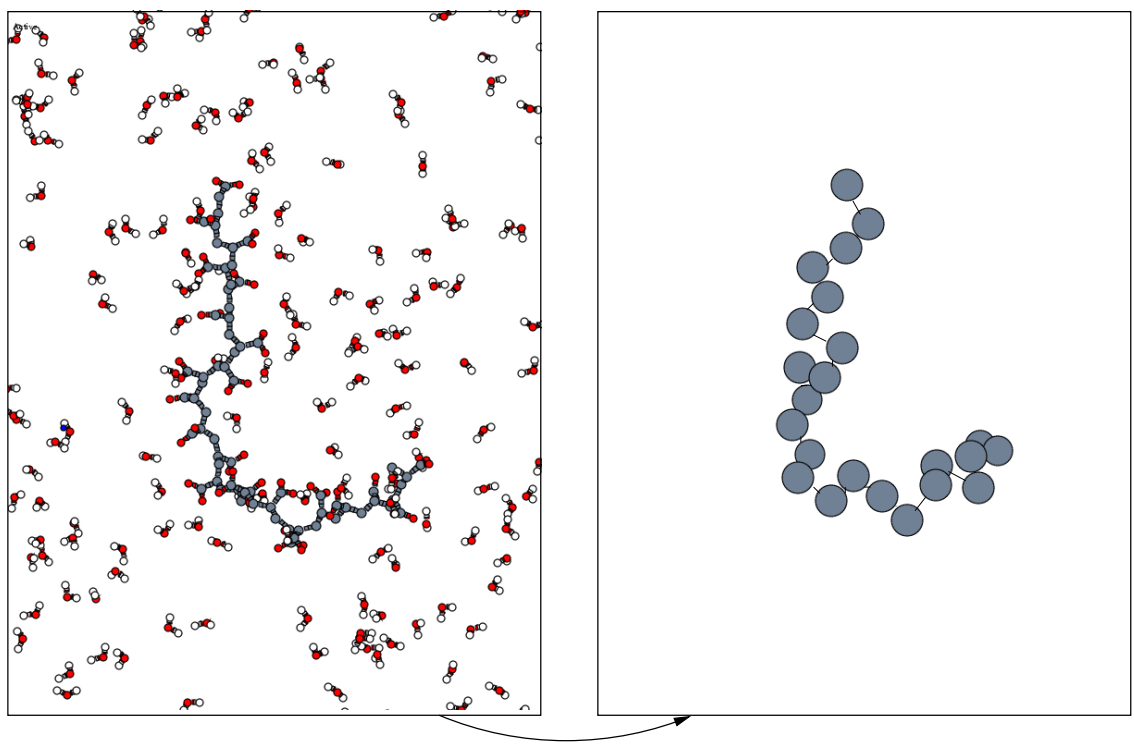}
  \end{center}
  \caption{Mapping from atomistic model to mesoscale model. The atomistic model
    comprises one poly (acrylic acid) (PAA) 23-mer, water and counterions 
    to achieve charge neutrality. The PAA chain with one bead per atomistic
    monomer is the only component reappearing after the mapping.}
  \label{at2cg}
\end{figure}

With our approach, we map the atomistic system with its detailed chemistry to
a coarser, mesoscopic model. This is done in continuous space. 
The crucial point is to incorporate the neglected atomistic degrees of freedom 
in some way into the 
coarse-grained model. The coarse-grained monomers are therefore designed
such that they can be identified with specific chemical groups of the polymer
itself (as an example, cf.\ Fig.~\ref{at2cg}). To gain the adequate ``local'' 
structure, we have to construct suitable intra-molecular potential terms.
For the ``global'' structure of a polymer, the non-bonded interactions play
a major role and, hence, they also have to be adjusted correctly. To do so, 
we first have to calculate the appropriate bond length, angular and 
torsional statistics
of the atomistic simulations. For the fitting procedure, we adapted our 
simplex optimization algorithm which was originally implemented for the 
development of atomistic force fields.\cite{faller99}
This procedure was already successfully applied for coarse graining protocols
of dense molecular liquids such as diphenyl carbonate or 
tetrahydrofurane.\cite{meyer00}

In this contribution, we extend the optimization against structural properties
to vinyl polymers. For diversity, we chose a melt of poly (vinyl alcohol)
(PVA) and an aqueous solution of poly (acrylic acid) (PAA) in form of its 
sodium salt. We show explicitly how to derive 
the coarse grained force fields starting from atomistic simulation data.
All parametrizations as well as the most important steps of the simplex
optimization are given. We focus on the interplay of various parameters
and discuss some technical problems we had to overcome in the course of 
the optimization process. A first comparison with experimental data  
validates finally the usefulness of the coarse-grained PAA force field. 

\section{Methods}

%
%
%
\subsection{Conceptional\label{concepts}}
Potentials for coarse-grained (CG) polymers may not only incorporate
energetic, i.e.\ local aspects of the underlying microscopic model. Due
to many-body effects, they also have to account for the entropic
contributions from the conformational degrees of freedom of the chain.
Therefore, we utilize structural information like intra- and inter-chain
distribution functions which contain the entropic information as target 
functions to construct our CG force field.

We start from the approach used by Tsch\"op et al.\ for the simulation of
polycarbonate (PC) melts.\cite{tschoep98a,tschoep98b} They assumed that
the distribution functions for length, angles and torsions taken between 
the coarse-grained beads directly
factorize into independent distribution functions of the above
variables.  This worked reasonably well: It was tested for three
different PC modifications. After remapping to an atomistic model, the 
resulting microscopic structure
compared well to neutron scattering data.\cite{eilhard99} The main
difference to our course of action is, that they derived their torsional
distributions via quantum chemistry calculations. Without any fitting
procedure, they then assumed the Boltzmann-inverted functions as
potentials for their force fields.

By contrast, we consider the construction of the force field as
a so-called inverse problem:
find an interparticle potential which reproduces a given
radial distribution function (RDF) or set of RDFs. The latter have proven 
to be suitable target functions in the case of simple liquids.\cite{meyer00} 
For polymer chains modifications originating from the connectivity are 
necessary: We use partial RDFs, the so called RDF-a and RDF-b, as
additional target functions. The first refers to an intrachain RDF for which 
we ignored the nearest neighbour chain particles. This can be easily 
done in computer simulations. The latter denotes the intrachain RDF 
excluding the two next neighbours.     

To perform an automatic optimization, we need three technical ingredients:
(i) specification of the potentials and parameters to be optimized,
(ii) evaluation of the quality of the total test potential, 
i.e.\ a merit function defining a hypersurface
in the parameter space which has to be minimized, and
(iii) rules to modify the test potential to find a better fit
of the target function.
While the latter two points are rather technical and easy to
implement, the specification of the functional form of the 
inter-particle potentials turns out to be the most difficult part.
For details on the simplex optimization scheme as we use it we refer
to Refs.\ \citen{faller99} and \citen{meyer00}.

Atomistic polymer force fields are usually decomposed into two main
parts: bonded and non-bonded potential. For both there are several
terms contributing which describe the energy required for distorting the
polymer in a specific fashion. In general, the total force field energy 
can then be written as follows. 
\begin{eqnarray}
  \label{eq:1}
  V_{tot} & = & V_{bonded} + V_{nb} \\\nonumber
          & = & (V_{str} + V_{bend} + V_{tors}) + (V_{vdw} + V_{es})
\end{eqnarray}
$V_{str}$ is the potential accounting or the bond stretching between two
atoms, $V_{bend}$ the energy needed for bending a bond angle, $V_{tors}$
represents the torsional energy, $V_{vdw}$ accounts for the excluded volume
repulsive as well as the London attractive forces between atoms 
and $V_{es}$ finally
describes the electrostatic interactions. The coarse-grained force field
shall be designed in the same way. However, having so many potential
terms in mind, it is not meaningful trying to optimize them all at the same
time. Instead, a more viable way to proceed is the successive adjustment of
the terms, in the order of their relative strength. 
\begin{equation}
  \label{eq:2}
  V_{str} \quad\rightarrow\quad V_{bend} \quad\rightarrow\quad %
  V_{vdw} \quad\rightarrow\quad V_{tors}
\end{equation}
Disregarding $V_{es}$, Scheme~(\ref{eq:2}) illustrates the relative rigidity 
of the force
field terms, assuming that $V_{vdw}$ is switched off for two adjacent atoms
next to an arbitrary atom in a polymer chain. Therefore, it makes no sense, 
e.g., to optimize $V_{vdw}$ if
$V_{str}$ has an inappropriate shape. We will, hence, start with the 
stretching energy, working our way systematically down to the torsional 
energy. 

The implementation is performed differently for PVA and PAA, respectively. We
present them in the subsequent Section~\ref{cgopt}.
Note that we do not consider electrostatic forces in our force fields. In the 
case of poly (vinyl alcohol), this is clearly justified because it represents
a neutral polymer. For the (fully) deprotonated poly (acrylic acid), one has
to be very careful.
Taking away the explicit treatment of the charges implies, that the 
force field we develop can not be applied to derive some specific properties
of PAA. Especially dynamical properties can not be expected to be calculated 
correctly. However, if concentrating on static properties and the fact that
we want to focus on the entropic contributions of the free energy, our
approach is still meaningful, as we will point out in 
Section~\ref{map&target}.

\subsection{Computational}
All simulations are performed in the NVT ensemble. The system consists of
an orthorhombic box employing periodic boundary conditions.  The Langevin 
equations of motion are integrated by the velocity Verlet 
algorithm with a time step $\Delta t = 0.01\tau$.\cite{All87} 
(Brownian Dynamics is assumed to be most 
appropriate for our mean-field type coarse-graining simulations.)
For technical reasons the potentials are normalized such 
that a temperature of $k_{B}T=1 $ is achieved. This temperature is 
maintained by the Langevin-thermostat with friction constant 
$\Gamma = 0.5 \tau^{-1}$.\cite{grest86}

Starting configurations are generated as random walks according to the given 
bond length, bond angle and torsion angle statistics (for the isolated chain).
The chains are then randomly placed into the simulation box, taking into
account periodic boundary conditions. The chains will now have some strong 
overlaps. To avoid excluded volume problems, the non-bonded potential is, 
hence, switched on slowly.
That allows the chains to relax the overlaps smoothly. 

We equilibrate our polymer systems as follows:
Either, we use a newly generated start configuration or, which represents 
the regular way, we take the last configuration of a previous run. 
A pre-equilibration procedure follows (a few thousend simulation steps 
with very short time interval). Here, we apply an artificial upper bound to
the forces to relax overlapping monomers in a well controlled way. 
This can be necessary for the system to adapt locally to the new simulation
parameters. Then, equilibration runs are performed. Here, we check some
thermodynamic standard properties (such as pressure or density) of the
system. We run the simulation until 
these properties are stabilized or the maximum number of equilibration runs is
exceeded. In the latter case, the setup will be abandoned. Otherwise, the
production run is started. Run lengths of $\approx 5\cdot10^5$ integration 
steps after equilibration turned out to be sufficient to determine 
smooth RDFs. The two examples treated in the following are at different 
densities: the PVA is in the dense melt, whereas the PAA is considered in very
dilute solution.

\section{Coarse graining of polymer oligomers}
\label{cgopt} 
\subsection{Poly (vinyl alcohol) oligomers}
The coarse graining of PVA is based on all-atom simulations with the
force field published in Ref.\ \citen{MPvG97poly} using the program YASP
\cite{yasp}.
The atomistic system containing 48 atactic decamers 
[~CH$_3$CHOH(CH$_2$CHOH)$_9$CH$_3$~] is simulated in an orthorombic box
with periodic boundary conditions at constant pressure (101.3 kPa) and
constant temperature (T = 500 K). The integration timestep is 2 fs
and the coupling times of the
Berendsen manostat and thermostat are 5.0 ps and 0.2 ps (respectively).
At T = 500 K, a trajectory of 10 ns  with frames every 2 ps
is generated of which we use the last 5 ns for evaluation of
target distributions. The average density is 1.03 kg/m$^3$.
Distributions of bond lengths, angles, and torsions are determined
excluding the outermost methyl groups.  For comparison, analysis of the
PVA trajectories of Ref.\ \citen{MP98jms} with the lowest water content
at T = 375 K yields quite similar distributions.

\subsubsection{Mapping and target functions\label{PVAmap}}
\begin{figure}
  \psfigure{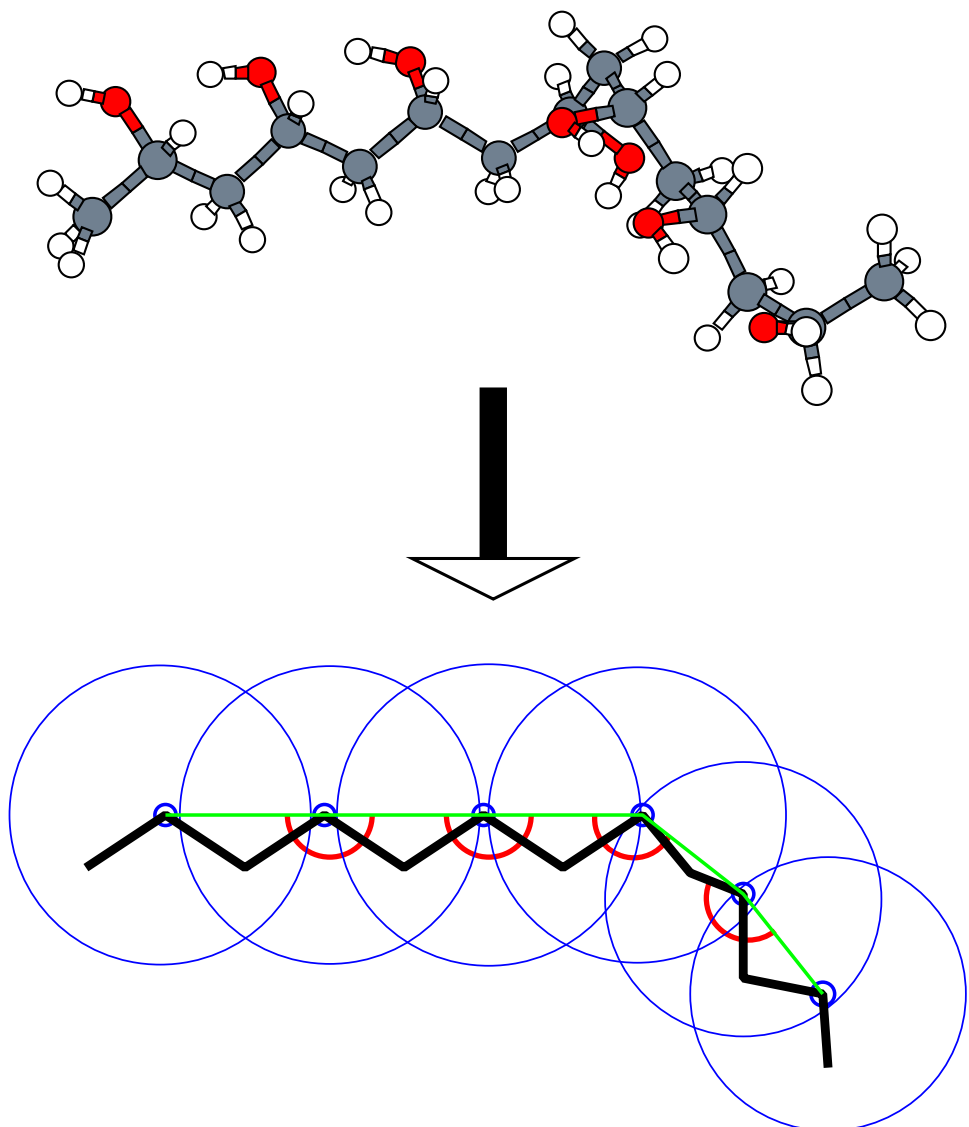} 
  \caption{Illustration of the mapping of the atomistic PVA oligomer
    to a coarse-grained bead-spring model. One bead represents one
    chemical repeat unit, a spring spans two carbon-carbon
    bonds on the backbone. The angle between three CG beads results from
    two successive atomistic torsions.}
  \label{f:pva}
\end{figure}

\begin{figure}
  \psfigure{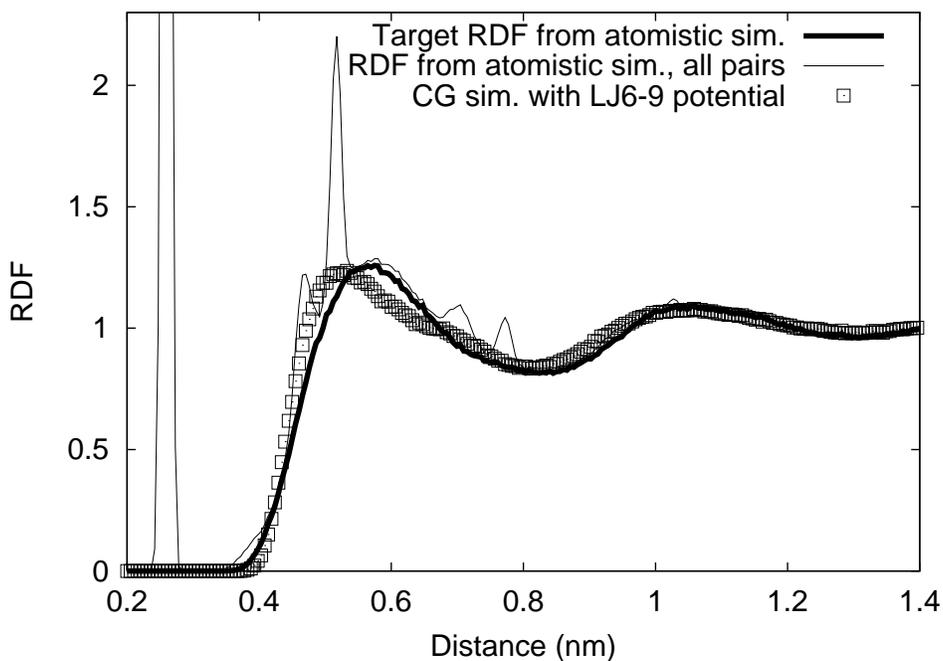}
  \caption{Model PVA monomers with a LJ6-9 potential.
    Thick continuous line: target distribution from atomistic simulations
    with 10mers (RDF between all monomers excluding the 4 nearest
    neighbors on each chain), thin line: RDF between all monomers from
    atomistic simulations (the sharp peaks at 0.26 nm and 0.52 nm are due to 
    the first and second nearest neighbors on the chains).  Squares: RDF
    of CG simulation with LJ6-9 potential. Best RDF out of 20 simplex
    steps. }
\label{f:pvardflj69}
\end{figure}

The first step of the coarse-graining procedure is to choose a
center of the CG beads. As we motivated in the introduction, we want to go 
beyond the simple united atom
model (incorporating hydrogens into their parent carbons) and
choose one CG bead per chemical repeat unit, i.e. two backbone
carbons.  Natural choices are (i) the center of mass of the monomer
or (ii) every second carbon atom on the backbone.  With respect to both 
mapping points, the bond length and angular distributions are
determined. The distributions look quite similar for both
possibilities. However, for choice (ii) they are slightly
sharper (not shown). Moreover, the bond length distribution is 
entirely independent
of the angular distribution. This is very convenient, since one has not
to worry about cross-correlations which would make the mapping more
complicated.  For this reason, we finally choose the backbone carbon of
the CHOH group as the mapping point for the CG beads
(see Fig.~\ref{f:pva}). As target RDF we use the distribution function where 
the distances between the first four neighbors along a chain are excluded.
The rationale is that the non-bonded potential of a dense melt should rather 
reflect the
interchain interaction. For the (short-range) intrachain interactions
there are already the bond stretching and bending potentials to determine
the right structure.
Fig.~\ref{f:pvardflj69} contains both, the RDF of all pairs as well
as the target RDF where the first four neighbors along the chain are excluded.
The target RDF is quite similar to typical RDF of a dense LJ-liquid.

\subsubsection{Potential parameterizations}
A Gaussian bell function $P(l)$ is found to represent the distribution of 
two successive backbone carbons of the CHOH group (not shown). 
It is Boltzmann-inverted to obtain $V_{str}$ as a harmonic potential, as
stated in Equation~\ref{eq:3}. For PVA, 
the bond length between neighboring CG beads is $0.26$ nm with
$\sigma^2=0.006$ nm$^{2}$ according the choice of the mapping.
\begin{equation}
\label{eq:3}
V_{\rm str}(l) = kT \ln\left( P(l) \right)
\end{equation}

\begin{figure}
  \psfigure{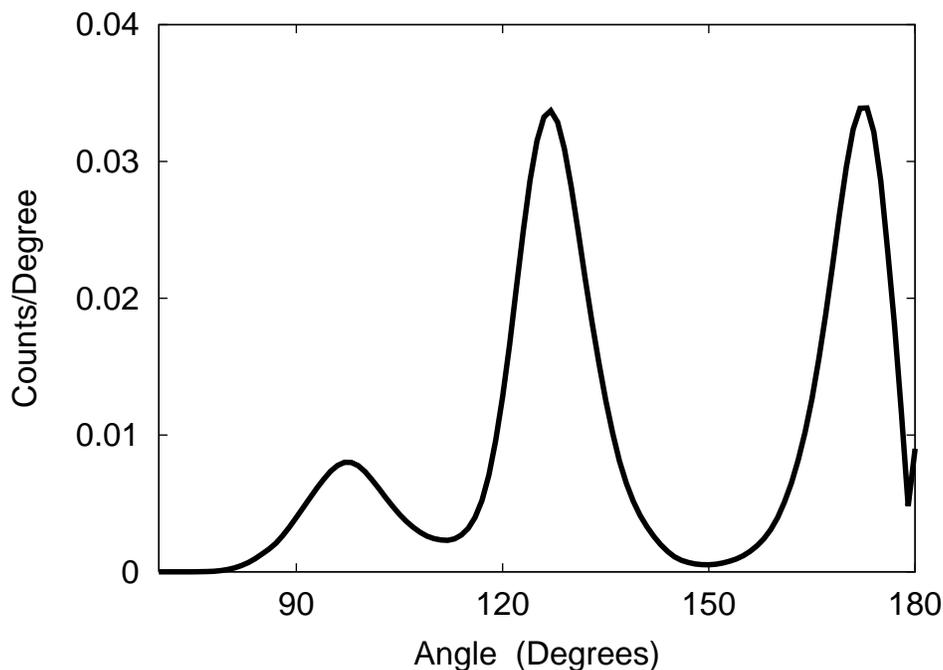}
\caption{Histogram of angles between successive coarse-graining points
  (every second backbone carbon) for PVA obtained from the atomistic 
  simulation.}
\label{f:pvaang}
\end{figure}

\begin{figure}
  \psfigure{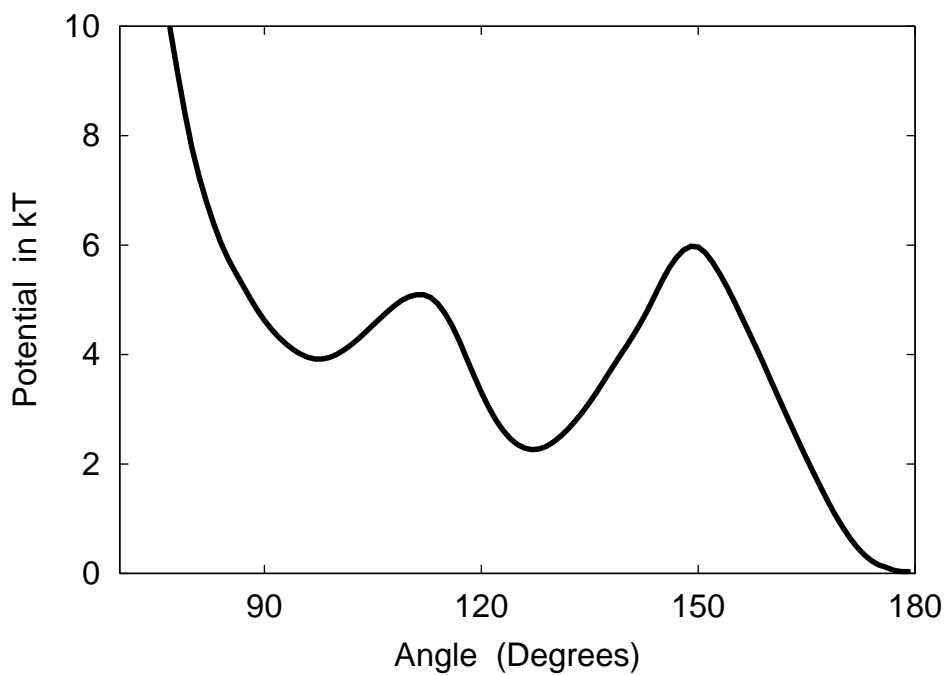}
\caption{Potential of mean force of the CG-angle distribution for PVA.}
\label{f:pvaangpot}
\end{figure}

We analyse the intrachain distributions of the atomistic trajectories
in order to find suitable CG potentials. 
The angular distribution between three successive C$^{(2)}$ atoms is
shown in Fig.~\ref{f:pvaang}. It exhibits 3 peaks which can be attributed to
trans-trans, trans-gauche and gauche$^+$-gauche$^-$ states of the two
dihedral angles on the atomistic backbone between the CG points.
The Boltzmann-inverted angular distribution (potential of mean force, 
cf.\ Fig.~\ref{f:pvaangpot})
\begin{equation}
\label{eq:Vbend}
V_{\rm bend}(\alpha) = kT \ln\left( P(\alpha)/\sin\alpha \right)
\end{equation}
is used as a first approximation of the angular potential in the CG
simulations.  In this case it turns out that the CG simulation yields almost 
the same distribution and, hence,  no further optimization needs to be 
performed. Since the torsional distribution between four successive CG points
does not show any particular structure, we run the CG simulations
without a torsional potential. 

The last issue of the CG force field is, thus, the non-bonded interaction 
which determines the excluded volume. The optimization of this non-bonded
potential according to the procedure of Ref.\ \citen{meyer00}
is described in the next subsection. The target RDF is quite `well-behaved' 
with a broad first peak (Fig.~\ref{f:pvardflj69}).
After the experience of Ref.\ \citen{meyer00} we need
a potential which is softer than a simple Lennard-Jones (LJ) type
potential and we start with the following general ansatz:
\begin{equation}\label{e:potraw}
  V_{vdw}(r)    =  {a_6 \over r^6} +{a_8 \over r^8} +{a_9 \over r^9} %
                         +{a_{10} \over r^{10}} + {a_{12} \over r^{12}} 
\end{equation}
Note that it is switched off 
for directly linked CG beads as well as for next nearest neighbours in 
order to seperate its influence from the bond angle potential.

%
%
%
\subsubsection{Optimization procedure}
The volume is fixed to obtain the correct density, whilst the non-bonded
potential is varied until the RDF of the corresponding atomistic
distribution is well reproduced.  The merit function which is minimized
is a least squares difference between the RDF $g_{\rm CG}(r)$ of the CG
system and the atomistic target RDF $g_t(r)$ integrated over a certain
range:
\begin{equation} \label{e:meritrdf}
f_{\rm rdf} = \int_{r_{\rm min}}^{r_{\rm max}} w(r) 
            \left(g_t(r) - g_{\rm CG}(r)\right)^2  dr \;.
\end{equation}
$w(r)=exp(-r)$ is a weighting function which additionally
penalizes deviations at small separation. \cite{meyer00}
Practically, the integration is performed as a sum with 
stepsize $dr=0.0052$ nm.

As the target RDF of PVA is similar to the RDF of dense LJ-liquids, we 
started with a LJ6-9 potential as a first approximation:
\begin{equation} \label{e:lj69}
V(r) = -\varepsilon \left(\left({\sigma\over r}\right)^6 
                       - \left({\sigma\over r}\right)^9 \right)\;.
\end{equation}
The LJ-type parameters $\varepsilon$ and $\sigma$ are optimized
with a two-dimensional simplex.
The cutoff radius for the CG-simulation is 1.1 nm (the same as in
the atomistic simulations) as well as the cutoff for integration in
the merit function.
The best trial RDF we find within 20 steps is shown in 
Fig.~\ref{f:pvardflj69}.
The corresponding potential parameters are $\varepsilon=0.5$ kT/mol
and $\sigma=0.53$ nm. The slope of the first peak is still too high. 
We expect no significant improvement with the potential of 
Equation \ref{e:lj69}. This is because the
ascent of the main peak has a smaller slope than it may be reproduced by
the LJ6-9 potential with, \textsl{at the same time as, } 
the second peak still fitting.

\begin{figure}
  \psfigure{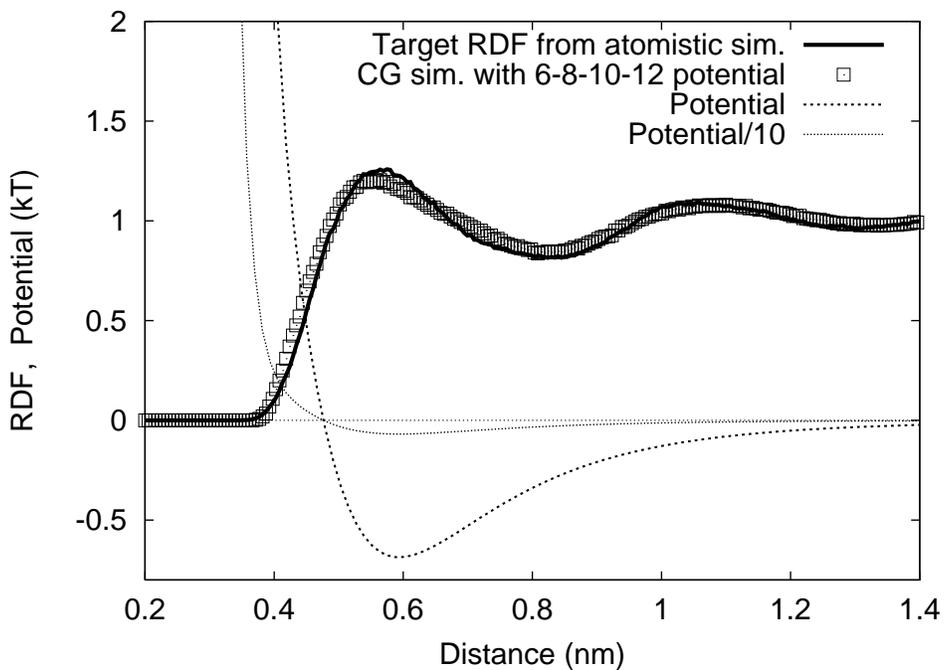}
\caption{Model PVA monomers with a 6-8-10-12 potential.
  Thick continuous line: target distribution from atomistic
  simulations with 10mers (RDF between all monomers excluding the 4
  nearest neighbors on each chain). Squares: RDF of CG simulation with
  the 6-8-10-12 potential. The corresponding potential is also plotted.}
\label{f:pvardf1}
\end{figure}

Next, potential (\ref{e:potraw}) is used with all even
powers in the following parametrization:
\begin{equation}\label{e:potmm}
V(r) = - {\varepsilon \over r^{12}} \left( r^{6}
         - 0.75 \, (\mu_{1}^{2}+\mu_{2}^{2}+\mu_{3}^{2}) \, r^{4}
         + 0.6  \, (\mu_{1}^{2}\mu_{2}^{2}+\mu_{2}^{2}\mu_{3}^{2}
                 +\mu_{1}^{2}\mu_{3}^{2}) \, r^{2} 
         - \nu \, \mu_{1}^{2}\mu_{2}^{2}\mu_{3}^{2}  \right)   \;,
\end{equation}
where $\varepsilon$ is a global scaling factor, $\mu_i$ are
the locations of the extrema and $\nu$ is a sort of
smoothing parameter (this potential is derived in
the appendix of Ref.\ \citen{meyer00}).
A 4-dimensional simplex is used with $\varepsilon$,
$\mu_1^2=\mu_2^2$, $\mu_3^2$, and $\nu$ as variables.
With $\nu$ slightly above 0.5 this yields a potential which resembles
a LJ-type function, however, the minimum being  much broader.
After guessing some start values for this parametrization,
the simplex algorithm was stopped after 32 steps.
The best result was quite similar to that shown in Fig.\
\ref{f:pvardf1}.

\begin{figure}
  \psfigure{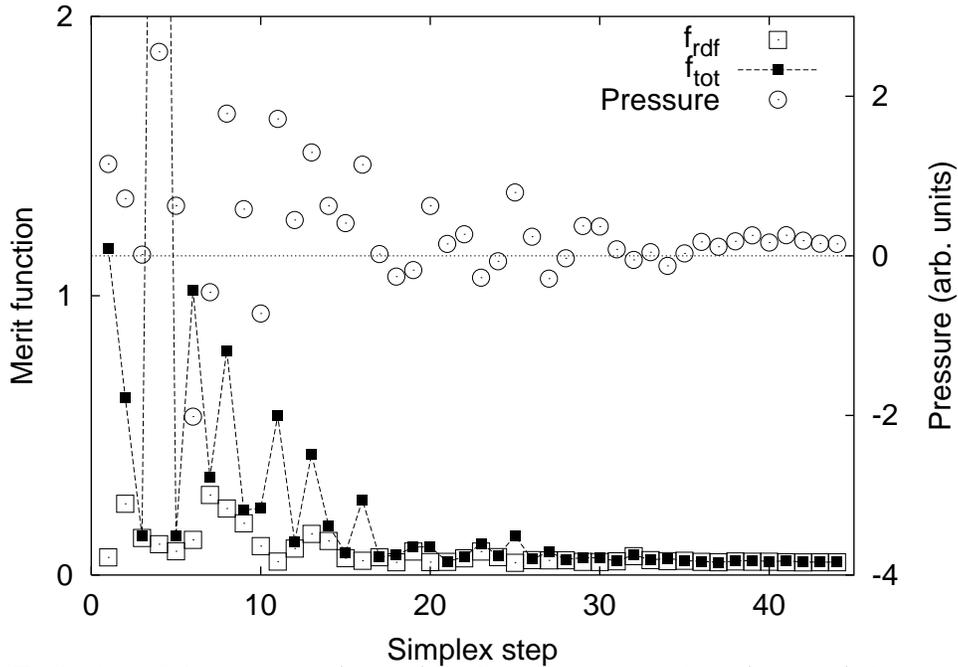}
\caption{Evolution of the pressure (circles) and the merit
  functions (squares) during one simplex optimization run for PVA
  10mers.  Open squares show the merit function $f_{\rm rdf}$ of the
  deviation from the target RDF. The filled squares represent the
  total merit function according to equation (\ref{e:meritpress}.)}
\label{f:pvaopt}
\end{figure}

However, with respect to future applications that will involve 
constant-pressure simulations, we are not satisfied just reproducing 
the RDF, but we would like to have the excluded volume in equilibrium
with the experimental density. This means we are searching
for a potential yielding a pressure near zero. For this,
we modify the merit function of the simplex algorithm
to include a term measuring the deviation of the pressure from
zero:
\begin{equation} \label{e:meritpress}
 f_{\rm tot} = f_{\rm rdf} + w_p (p-p_0)^2 \;,
\end{equation}
with the target pressure $p_0=0.1$ (LJ units) 
and a relative weight $w_p=0.2$.
This gives still a rather small weight to the pressure to permit
sufficient fluctuations for testing the parameter space.
The evolution of the pressure and the merit function
is shown in Fig.~\ref{f:pvaopt}). The algorithm converges to
a minimum of the merit function which represents a satisfying result.
The best value is already obtained in step 21 and is the one shown
in Fig.~\ref{f:pvardf1}.
The corresponding parameters are $\varepsilon=12$ kT,
$\mu_1=0.41$ nm, $\mu_3=0.58$ nm, and $\nu=0.517$.
 Fig.~\ref{f:pvardf1} contains also the  potential scaled
by a factor of 10 to show its two repulsive
regimes: a really excluded hard shell below 0.4 nm and
a softer core above this value to the minimum at 0.59 nm.  

%
%
\subsection{Poly (acrylic acid) oligomers}
%
%
For PAA, the optimization is based on atomistic simulation data 
(also generated using the YASP package) by Biermann and M\"uller-Plathe
\cite{biermann00}. They studied one fully deprotonated,
atactic oligomer strand of 23 monomers in 3684 water molecules at ambient 
conditions. 
Sodium counterions were included to achieve overall charge neutrality. 
The total simulated time was $9$ ns. We map this system to the 
mesoscale by replacing one repeating unit (i.e.\ one monomer) by 
one bead. Additionally, the water solvent is removed as shown in 
Fig.~\ref{at2cg}. The simulated CG system contained 460 particles (20 
23-mers, for technical reasons) at density $\rho =0.1962\sigma ^{-3}$. 
This density is adapted from the atomistic simulation and corresponds to the 
highly diluted regime. The chains do practically not interact with each other.
Moreover, one must be aware of the problem that we are dealing with a highly 
charged polymer. One might think, that the long-range character of the Coulomb
interactions does not allow for them to be neglected in the CG model. In our 
case this is not correct for
two reasons: First and most importantly, the Debye length (characterizing the
relative importance of the electrostatic energy compared to thermal energy)
of the system is close to the Debye length of water, which is $0.7$ nm. 
This corresponds approximately to the distance of two non-bonded CG monomers. 
Hence, the Coulomb interactions can be neglected for larger distances to a
good approximation. 
Second, the atomistic simulations were (for sake of the first reason)
performed with a reaction field approximation of the Coulomb interactions,
implying that already the atomistic simulation does technically not include 
any long-ranged interaction.
Note finally, that for short distances charges will still be present 
implicitly in the CG force field. They are, as the PAA-solvent interactions, 
reflected in the intrachain RDF and in the CG potential resulting from it. 

\subsubsection{Mapping and target functions\label{map&target}}
\begin{figure}[hb]
\begin{center}
\epsfxsize 8cm \epsfbox {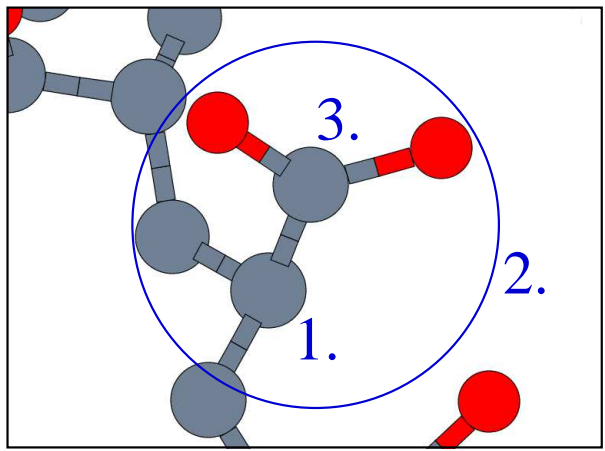}
\end{center}
\caption{Various possibilities how to define the sphere center of the coarse
  grained beads for PAA. Assuming that one bead should correspond to one
  repeating unit (-CH(-COOH)-CH$_2$), there are three obvious choices:
  (1) the CH backbone carbon atom, (2) the center-of-mass of the
  repeating unit and (3) the sidegroup carbon atom.}
\label{cg-bead-def}
\end{figure}

\begin{figure}[hb]
\begin{center}
\epsfxsize 10cm \epsfbox {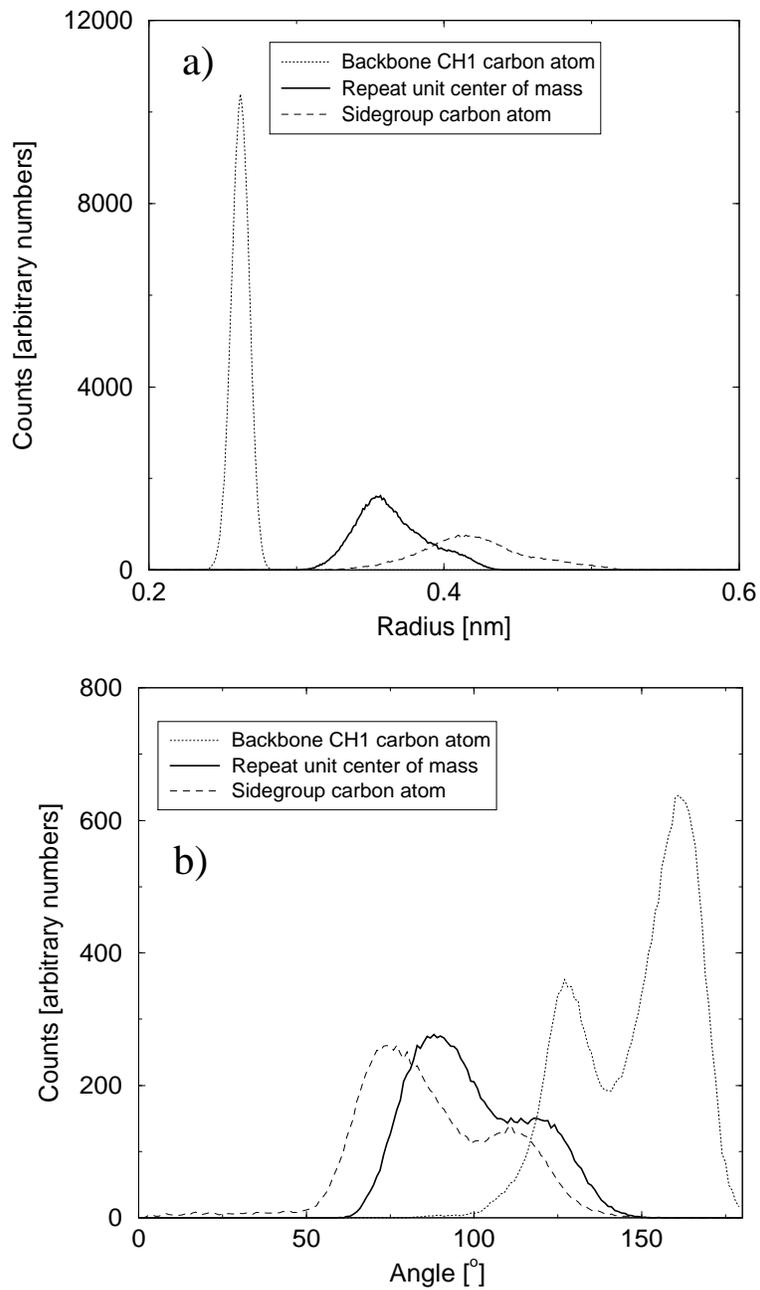}
\end{center}
\caption{Intrachain distance (a) and angle distributions (b) for three
  choices of coarse-graining points in the atomistic model of PAA, cf. 
  Fig.~\ref{cg-bead-def}.}
\label{at-distrs}
\end{figure}

Again, the first question is how to define the coarse-grained beads. Exactly 
as in the case of PVA, we want to incorporate an essential amount of
atomistic information in our coarse-grained model. For PAA, three 
choices seem to be appealing, cf.\ Fig.~\ref{cg-bead-def}. Either, we could
choose the backbone carbon which is linked to the side group (marked 1 in 
Fig.~\ref{cg-bead-def}) or the side-group carbon (2) or the center of mass 
of a (-CH(-COOH)-CH$_2$) repeating unit (3).
To make a decision, we revisited the atomistic trajectory and 
investigated the intrachain distance and
angular distributions for these candidates. They are presented in
Fig.~\ref{at-distrs}. For the distance distribution 
(Fig.~\ref{at-distrs}a), the sharpest peak originates from the backbone CH
carbon atom [possibility 1]. This is understandable because adjacent
CH carbons are directly linked via two covalent bonds, leaving not
much configurational freedom. In contrary, the sidegroup has rotational
freedom to move such that the distribution of the COO$^{-}$ 
carbon [possibility 3] is much broader around a much larger mean value. The 
center of mass of a repeat unit [possibility 2] lies somewhere in between. 
For the angular distributions, the first observation is that side group 
(3) and backbone carbon (1) distributions 
have very different peaks. 
In view of this, the behaviour of the center-of-mass distribution made us 
choose it for the mapping: it has a reasonably sharp distance peak and 
the angular distribution includes some of the behaviour of the side group. 
So, it represents a compromise between our preference for sharp peaks and
the need of taking into account the side group.  

\begin{figure}[hb]
\begin{center}
\epsfxsize 12cm \epsfbox {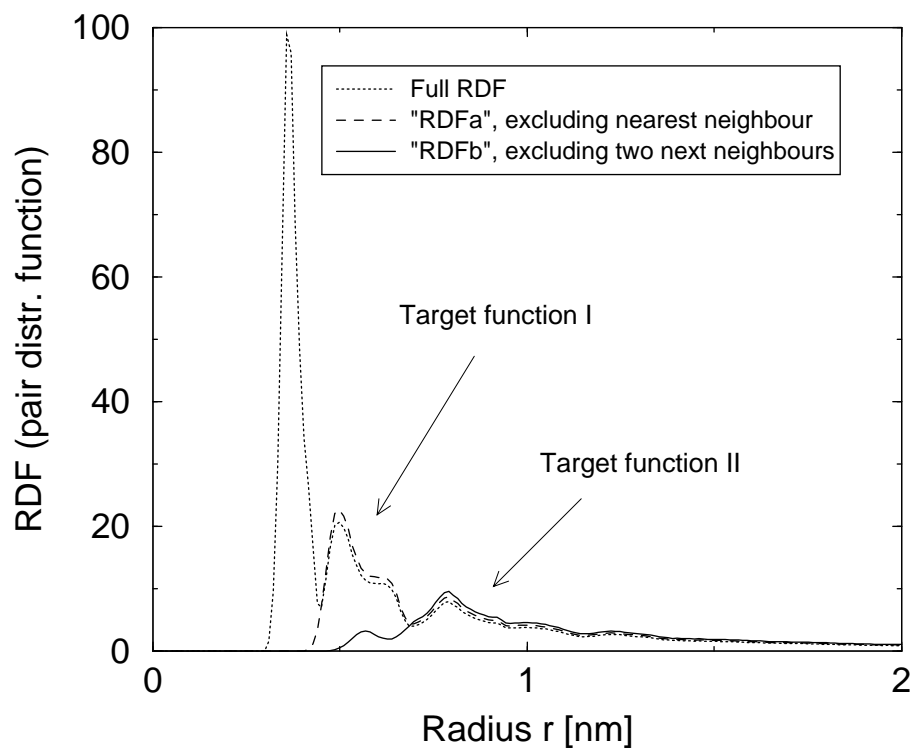}
\end{center}
\caption{Intrachain radial distribution functions of the repeating units
  center of mass of PAA. The 
  first peak originates from the bond length distribution, the next two peaks
  from the bond angle distribution. After this the influence of the 
  intrachain non-bonded interactions starts.}
\label{atomistic-rdfs}
\end{figure}

Now that we determined the mapping, we need a meaningful target function.
In contrast to PVA, exclusively intrachain properties can be chosen because
Biermann´s simulation contained only one polymer strand in solution. 
Analysis of the various atomistic intrachain radial distribution function 
(RDF, RDF-a and RDF-b, Section~\ref{concepts}) reveals that the full RDF can 
be decomposed (Fig.~\ref{atomistic-rdfs}). The first peak 
originates exclusively from the nearest neighbours as it would be the
only one which remains when taking the difference between full RDF and RDF-a.
So, this peak is determined by the distance distribution of adjacent 
center-of-mass points. Similarly, the next two peaks are mainly resulting 
from the second-nearest neighbours.
They correspond exactly to the peaks of the angular distribution, as shown
in Fig.~\ref{at-distrs} and, hence, are dominated by this distribution.
After that, non-bonded interactions start to influence in the RDF. 

For the
optimization procedure (and the target function(s)), it has the following 
consequences: as we stated in Section~\ref{concepts} the relative strength 
of the various potential terms should be taken into account. The strongest 
one is the bond stretching potential $V_{str}$. Since the 
corresponding peak is the overall sharpest and almost normally distributed, 
we skip any optimization in
favour of a reduction of parameters. Instead, we approximate the 
atomistic center-of-mass distance distribution by a Gaussian function 
$P(l)$ and Boltzmann-invert it to obtain $V_{str}(l)$ (cf.\ 
Equation~\ref{eq:3}). 
On the other side, the weakest potential we could take into account is the 
torsional potential. Although the distribution (not shown) is not flat as 
in the case of PVA, the question arises whether it is too unimportant to make
any significant contribution to the outer peaks in the RDF or not. 
In a first optimization process, we therefore decided to not include it in 
the coarse-grained force field at all. We discuss this point in more detail
below (Section~\ref{improvements}). 

So for the moment, we are left with the bond-bending potential $V_{bend}$ and 
the non-bonded potential $V_{vdw}$.
The optimization of these can be split up into three parts: First, $V_{bend}$ 
will have its strongest influence on the following
two peaks and is, hence,  optimized against the RDF-a. Second, the non-bonded 
potential has to be optimized. In order to separate its influence on the 
RDF-a main peaks, we switch $V_{vdw}$ off for 1-2 and 1-3
interactions. It is then optimized against the RDF-b.
It makes no sense to choose the full RDF or RDF-a for this
optimization because they are dominated by the 1-2 and 1-3 interactions, 
while the RDF-b reflects only 1-4 and higher interactions.
Third, the best parameters from the separate optimizations are taken as start
values for a mixed optimization against the RDF-a which responds strongest to 
variations of both the bond bending and non-bonded potential. 

\subsubsection{Potential parameterizations \label{paa_pot_paras}}
\begin{figure}[hb]
\begin{center}
\epsfxsize 12cm \epsfbox {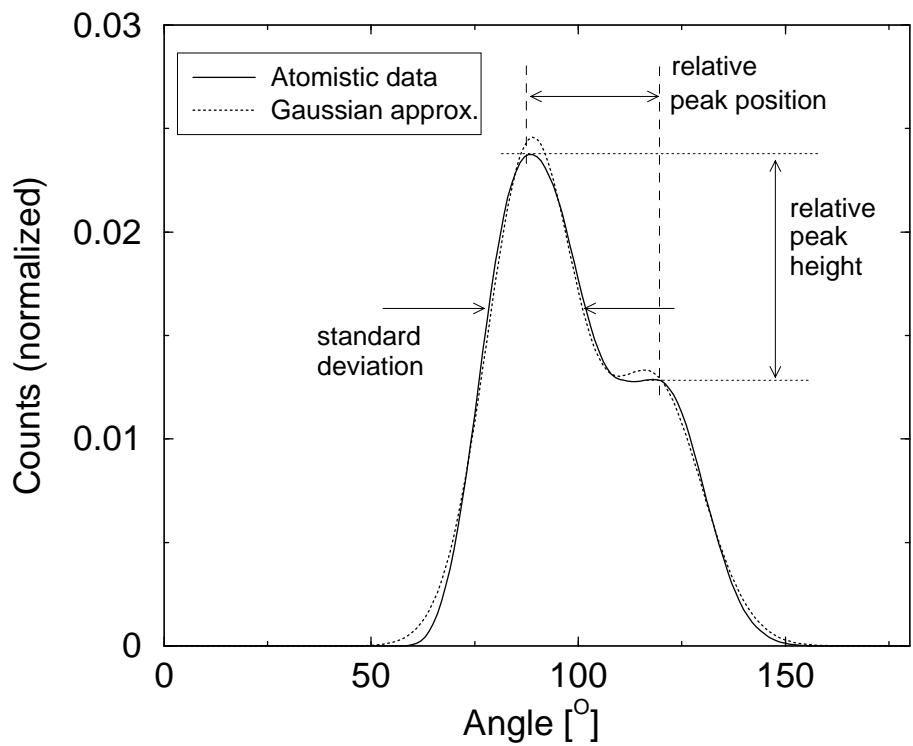}
\end{center}
\caption{Bond angle parametrization for the PAA CG model. The atomistic data 
  could be fitted by two Gaussians. This leaves four parameters for the coarse
  graining optimization: relative peak height, relative peak position and the
  two standard deviations.}
\label{bonded-parms}
\end{figure}

%
%
%
For the CG bond streching potential, the atomistic output distribution of two
successive center-of-mass monomers is investigated. The mean 
value of the Gaussian $P(l)$ is determined to be $0.363$ nm and the standard 
deviation is $0.022$ nm. These values are neither changed nor optimized in
the CG model, as stated in Section~\ref{map&target}, but transformed into a
potential energy according to Equation~\ref{eq:3}. The atomistic output 
distribution for the center-of-mass angles can be approximated well by two 
Gaussians, as shown in Fig.~\ref{bonded-parms}. In contrast to the PVA model,
we do not convert \textsl{this} distribution into a potential function. 
Instead, we use an auxiliary distribution which, transformed, gives the CG
bending angle potential. It has the same qualitative shape but differs in
details. The approximation provides us 
with four parameters for the CG optimization: relative peak 
height, relative peak position and the two standard deviations. The values 
of the fit are given in Table~\ref{table-bond-parms}. The standard deviations 
of the two peaks are very similar. In order to lower the number of 
parameters, we sometimes freeze their ratio and constrain them to their
atomistic value of $\sigma^{G}_{1}/\sigma^{G}_{2} = 0.92$. Varying the Gaussian 
parameters, we can generate physically understandable new bond angle
distributions. They are transformed into CG bond angle potentials
$V_{bend}$, by application of Equation~\ref{eq:Vbend}.

For the non-bonded interactions, we perform simulations 
using a standard LJ6-12 potential with $\varepsilon$ and $\sigma$ as test 
potential to get a feeling for the excluded volume behaviour of the chains, 
i.e.\ $V(r)$ of Equation~(\ref{eq:potlj}) is applied. 
\begin{equation}\label{eq:potlj}
V_{\rm LJ6-12}(r) = 4\varepsilon \left( \left({\sigma\over r}\right)^{12} -
                 \left({\sigma\over r}\right)^{6} \right)  \;,
\end{equation}

A cutoff distance 
of $2.52$ nm is imposed for this potential. As for PVA, it is switched off 
for directly linked CG beads as well as for next nearest neighbours. 
Meaningful values for the LJ parameter are $\varepsilon = 0.384$ kT/mol 
and $\sigma = 0.74$ nm (no data shown).
However, the LJ potential is too inflexible to cope with the demands of a
suitable non-bonded potential for PAA. In particular, it does not reproduce
the correct rise and peak broadness of the target RDF(s). So, we need a 
specially designed non-bonded potential to succeed in our optimization task:
\begin{equation}
\label{e:potpiece}
V_{vdw}(r) = \left\{\begin{array}{ll}
          \varepsilon_1\,\left( \left({\sigma_1 \over r}\right)^8 
                             - \left({\sigma_1 \over r}\right)^6 \right) 
            &  r < \sigma_1 \\
          \varepsilon_2\,\left(\sin{(\sigma_1-r) \pi \over (\sigma_2-\sigma_1)2 }
                       \right)
            &  \sigma_1 \le r < \sigma_2 \\
          \varepsilon_3\,\left(\cos{(r-\sigma_2) \pi \over \sigma_3-\sigma_2 }
                      - 1 \right) - \varepsilon_2
            &  \sigma_2 \le r < \sigma_3 \\
          \varepsilon_4\,\left(-\cos{(r-\sigma_3) \pi \over \sigma_4-\sigma_3 }
                      - 1 \right) - \varepsilon_2
            &  \sigma_3 \le r < \sigma_4\equiv r_{\rm cut} \\
       \end{array} \right.
\end{equation}
This potential was previously applied to tetrahydrofurane (THF) and diphenyl 
carbonate
(DPC).\cite{meyer00} It consists of a hard core repulsion, using a LJ6-8
potential. This means it is weaker than a standard LJ6-12 which is useful
for CG simulations because center-of-mass points can approach each other
more closely than atomic cores for which the original LJ6-12 was designed.
However, a strictly forbidden excluded volume region remains.
The next two parts are also repulsive, with force minima at $\sigma_2$ and
$\sigma_3$. This enables us to generate an RDF with a first peak or a 
shoulder at $\sigma_2$ and a second one above $\sigma_3$. These features
are essential, as we will see in the next section.

\subsubsection{Optimization procedure}
The simplex optimization was discussed in full detail in foregoing 
publications.\cite{faller99,meyer00} 
Therefore, we will only state our start simplices and final results.

During the first stage, we optimize the CG bond angles against the RDF-a.
The non-bonded interactions are held fixed and simulated with some few test 
guesses for the LJ6-12 potential ($\varepsilon = 0.2 - 1.75$ kT/mol and 
$\sigma = 0.65 - 0.8$ nm). 
As stated in the previous subsection, they are not switched on before
the 1-4 distance between the beads. We, therefore, do not expect them to
have much influence on the 1-3 peaks which dominate the RDF-a. Therefore,
non-bonded test guesses are sufficient at this stage of the simplex 
optimization. As start simplex of a typical simulation at this stage, 
we use e.g.:
\begin{verbatim}
#        eps       sig     amp1/amp2   std1     merit function f_a
#1      0.384      0.74      1.95      5.0      234.644
#2      0.384      0.74      2.20      6.5      188.777
#3      0.384      0.74      1.90      7.5      251.345
#4      0.384      0.74      1.95      8.0      255.425
#5      0.384      0.74      2.50      9.0      208.890
\end{verbatim}

\begin{figure}[hb]
\begin{center}
\epsfxsize 10cm \epsfbox {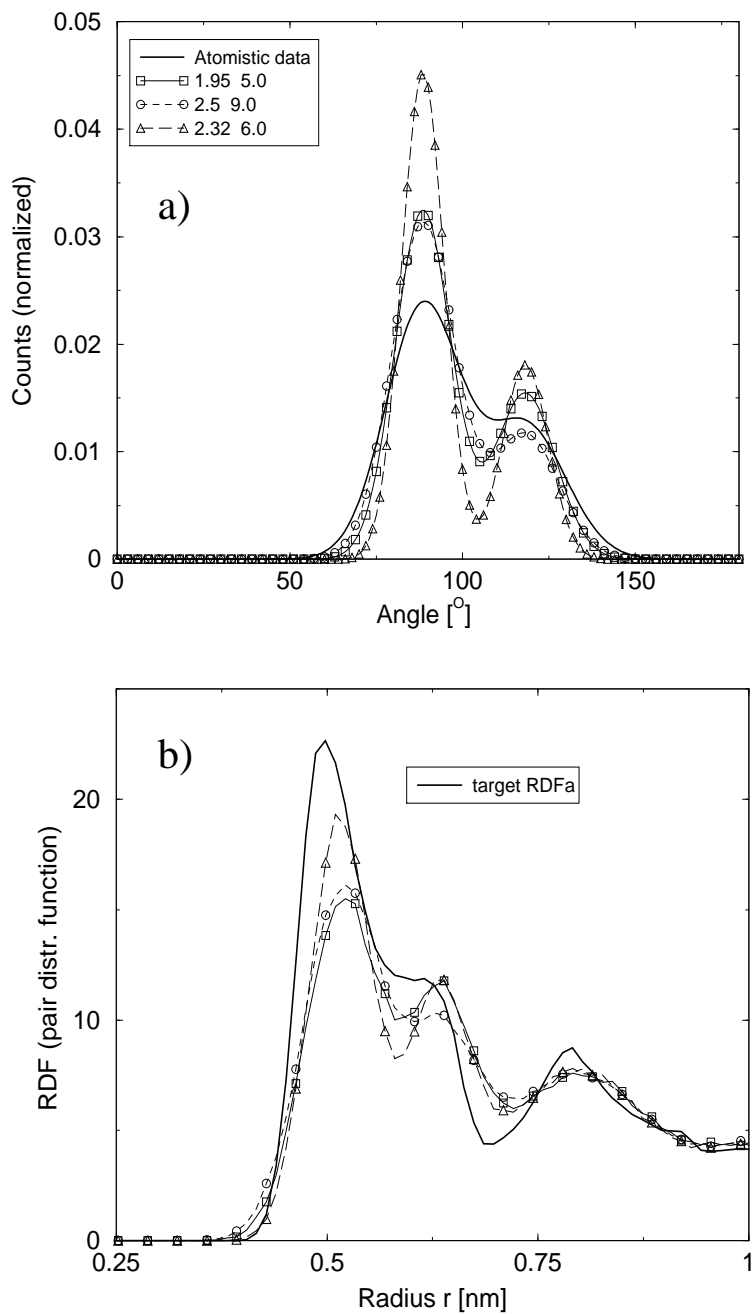}
\end{center}
\caption{Bond angle potential optimization against the RDF-a for PAA. 
  (a) Input distributions. The relative
  amplitudes (first number) and the standard deviation (second number) 
  of the main peak are optimized. (b) The resulting RDF-a. For the non-bonded
  interactions, a LJ 6-12 potential with $\varepsilon = 0.384$ and 
  $\sigma = 0.74$ is applied.}
\label{bonded-only-opt}
\end{figure}

Fig.~\ref{bonded-only-opt} shows some results obtained during the course
of the optimization
process. The relative amplitudes and the standard deviation of the 
distribution of the first 
peak are used as optimization variables (Fig.~\ref{bonded-only-opt}a). 
The standard deviation of the other peak is adjusted as described in 
Section~\ref{paa_pot_paras}. A qualitative match of the CG peaks and the 
atomistic peaks can be observed for the resulting RDF-a
(Fig.~\ref{bonded-only-opt}b). Still, the height of the main peaks is not 
sufficiently well reproduced and for too low a standard deviation ($< 6^{o}$), 
an unwanted dip appears between the main peaks (at $r\approx 0.6$ nm). It turns
out that it is favourable for the simplex to overshoot the intensity of the 
$90$ degree peak by $15-30 \%$ and to sharpen the peaks by a considerable 
amount of $30-50 \%$ compared to the atomistic initial distribution. This 
follows from the fact, that the weight function decays exponentionally. 
Therefore, inner peaks are tried to be matched first. For the sharpening
of the CG input distribution, however, there is also a physical reason:
the target pair distribution functions we observe in the atomistic 
simulations is the Boltzmann-inverse of a potential of mean force (at finite
temperature). Hence, it is less sharp than the CG distribution which is the
Boltzmann-inverse of a potential energy at $T=0$ K. 

After some series of simulations it became clear, that the details could not 
be optimized by varying the bond angle parameters only. The relative width 
of both standard deviations need not be systematically modified with respect 
to the atomistic case. The same holds for the position of the main 
peak. This leaves us with three parameters for post-optimization at stage
three: side peak position, relative peak height and one standard deviation. 

\begin{figure}[hb]
\begin{center}
\epsfxsize 10cm \epsfbox {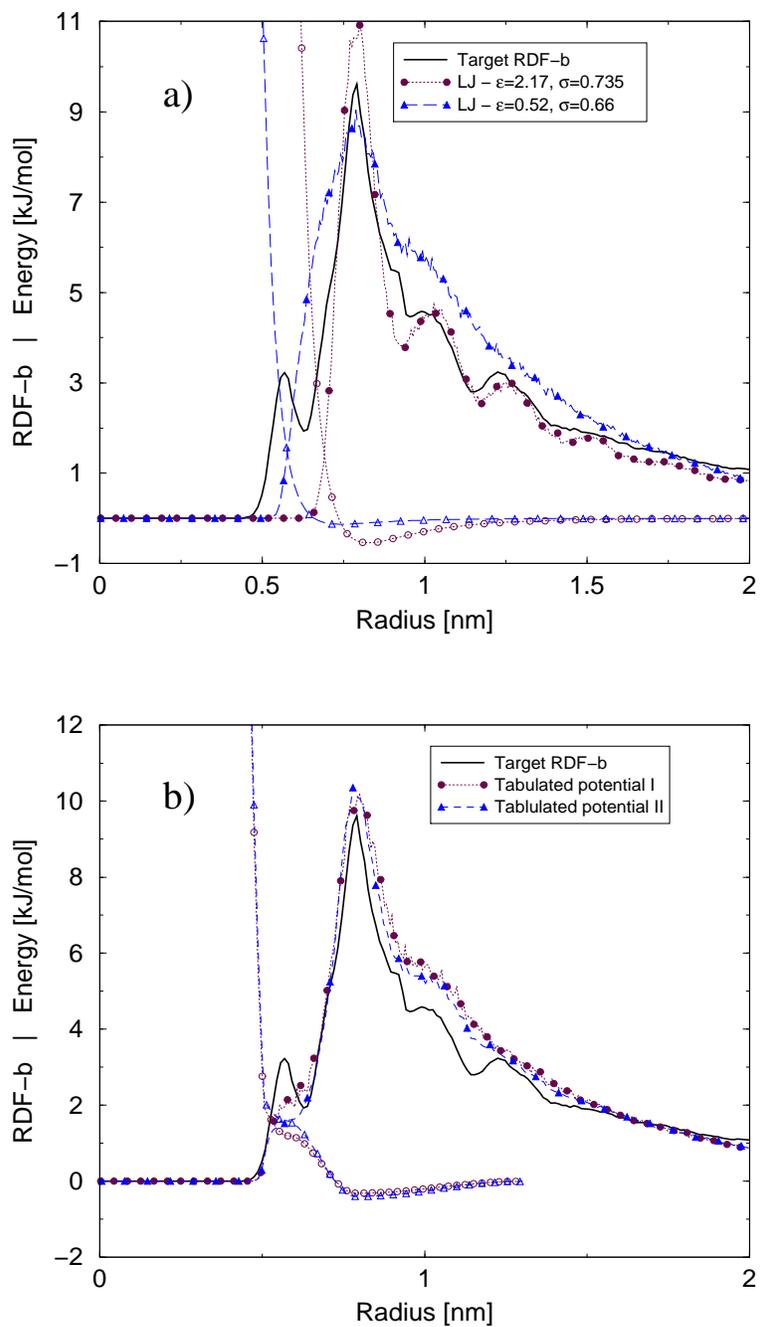}
\end{center}
\caption{Non-bonded potential optimization against RDF-b for PAA. 
  Open symbols denote
  the potentials, filled symbols the corresponding RDFs. Small changes in the
  potentials manifest themselves strongly in the trial RDFs. (a) 
  Lennard-Jones 6-12 potential of Equation~\ref{eq:potlj} (b) Piecewise 
  potential of Equation~\ref{e:potpiece}}
\label{nb-only-opt}
\end{figure}

Therefore, we fix an intermediate set and turn to stage two, 
the non-bonded parameter optimization against 
the RDF-b. Typical examples of the results are presented in 
Fig.~\ref{nb-only-opt}. Clearly, the LJ6-12 optimization is rather
unsuccessful (cf.\ Fig.~\ref{nb-only-opt}(a)). Only one of either the rise of 
the pre-peak or the rise of the main peak can be matched. The descent does 
not show at all 
the characteristic structure of the target function from the atomistic 
simulations. Therefore, we need to move over to a potential which is 
tailored to the following requirements: (i) a repulsive core to suppress
too close contacts and to mimic the sharp rise of the pre-peak correctly.
(ii) a tiny plateau to determine the width of the pre-peak. (iii) a second
softer
repulsive region to grasp the ascent of the main peak and (iv) an attractive
tail to adequately reproduce the main peak descent. (As it turns out, a 
well-shaped main peak of the RDF-b will be followed by acceptable side 
peaks.) The potential $V_{vdw}$ given in Equation~\ref{e:potpiece} fulfills 
all these demands. 

Test runs with this modified potential are already much better
compared to the LJ6-12 potential. However, $V_{vdw}$ consists of eight
parameters. This makes it difficult to optimize all of them at the same time
for two reasons. First, the computational demand gets quite high. Second, and
more importantly, such a simplex will be trapped more easily in
a local minimum, if one cannot guess at least some of the
parameters well. If this high-dimensional space is dominated by one or two
of the eight parameters (a very frequent case from our experience, cf.\ 
Ref.\ \citen{meyer00}), the merit
function will not change much if some weak parameters are driven in the 
wrong direction of a local minimum. Hence, we only choose three to five
parameters for one optimization series and fix part of them for a 
subsequent series in favour of optimizing another subset. 
Fig.~\ref{nb-only-opt}(b) shows two examples of a simplex started with:
\begin{verbatim}
#       sig1     sig2     eps2     eps3     eps4     merit function f_b
#1      0.505    0.065    0.5      2.5      0.3      44.1712
#2      0.510    0.065    0.6      2.0      0.4      17.8984
#3      0.520    0.053    0.4      2.3      0.5      23.8493
#4      0.505    0.060    0.5      2.1      0.5      19.0945
#5      0.505    0.068    0.6      2.4      0.4      34.4716
#6      0.510    0.060    0.4      2.2      0.6      24.4090
\end{verbatim}
Here, we fix the pre-optimized non-bonded parameters $\varepsilon_{1}=14.0$ 
kT/mol , $\sigma_{3} = 0.79$ nm and $\sigma_{4} = 1.3$ nm. In the case of 
$\sigma_{3}$, this 
is physically motivated: it corresponds to the potential minimum and matches
the main peak position of the RDF-b, for $\sigma_{4}$ the value is almost
arbitrarily chosen in the vicinity of values for which the LJ6-12 test
potential decayed to $0.01\cdot\varepsilon_{LJ}$. As best fit (shown in 
Fig.~\ref{nb-only-opt}), this simplex yields
\begin{verbatim}
#19     0.505137 0.073784 0.750695 1.512260 0.314712 10.9107
\end{verbatim}

It turns out that some parameters could be determined easily and precisely
individually while the interplay of others generates strong oscillations in
the merit function $f$. Those are $\varepsilon_{2} - \varepsilon_{4}$
because they massively influence the probabilities of the closest approach
of the CG beads, i.e.\ determine the relative strength of the peaks. 
Consequently, they are taken to be post-optimized in the third stage 
together with the crucial bond bending parameters. 

\begin{figure}[hb]
\begin{center}
\epsfxsize 8cm \epsfbox {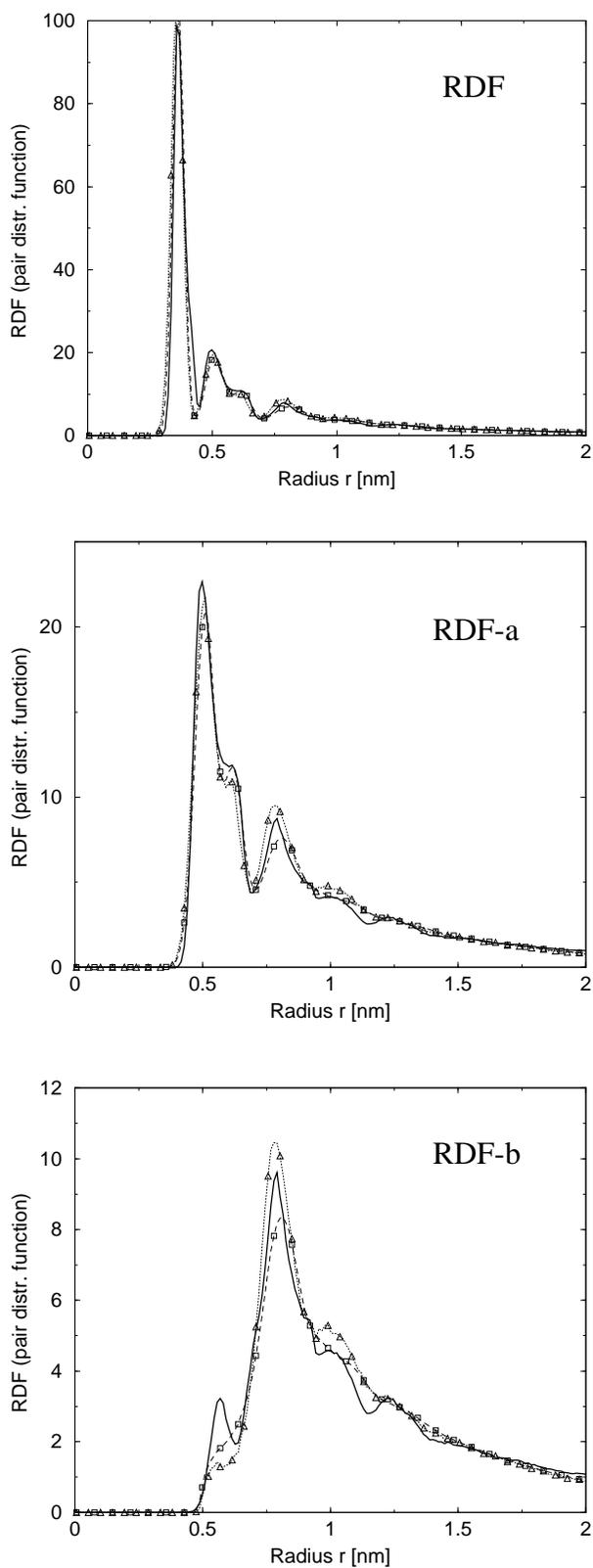}
\end{center}
\caption{Final optimization result for PAA. Thick lines: target RDFs from 
  atomistic
  simulations. Triangles: first optimization force field. Squares: improved
  force field, including torsions. All features of the RDF are at least
  qualitatively well reproduced.}
\label{best-opt}
\end{figure}

Some of the simplex points encountered during the final optimization runs
are the following:
\begin{verbatim}
#     amp1/amp2   peak2    std1     eps2     eps3     eps4    merit f_a
#2      1.9      (118.2)  (6.8)   0.700000 2.300000 0.350000  83.2977
#7      2.088    (118.2)  (6.8)   0.430000 2.548000 0.540000  50.2795
#8      1.983    (118.2)  (6.8)   0.372000 1.847200 0.566000  85.8116
.
.
#13   (1.90)    (118.2)  (6.8)   0.387200 2.304243 0.522016  48.9554
.
.
#7    (1.90)     117.031   6.743750 (0.39)  (2.31)  0.437500  47.1609
#24   (1.90)     116.060   6.716646 (0.39)  (2.31)  0.428761  41.0346
#33   (1.90)     116.555   6.769754 (0.39)  (2.31)  0.463141  39.4119
#42   (1.90)     116.709   6.768629 (0.39)  (2.31)  0.463242  37.843
\end{verbatim}
Numbers in brackets denote parameters which are not optimized in the
corresponding run. Although this is just a small selection from the 
optimization, the points demonstrate the
relative strength of the parameters. The first line contains 
(as it turns out) three almost optimized parameters, but the wrong
relative strength of $\varepsilon_{2}$ and $\varepsilon_{4}$ destroys
a good value of $f$. The latter is the overall most dominant parameter
in this line. The final result is listed in Table~\ref{table-bond-parms}
(bonded parameters) and Table~\ref{opt-nb-parms} (non-bonded parameters). 
It is visualized for all RDFs in Fig.~\ref{best-opt}. All functions 
are reproduced qualitatively correct. For such a complicated molecule
like PAA, this is all we could expect for a first trial. 

\subsubsection{An improved model 
  \label{improvements}}
\begin{figure}[hb]
\begin{center}
\epsfxsize 12cm \epsfbox {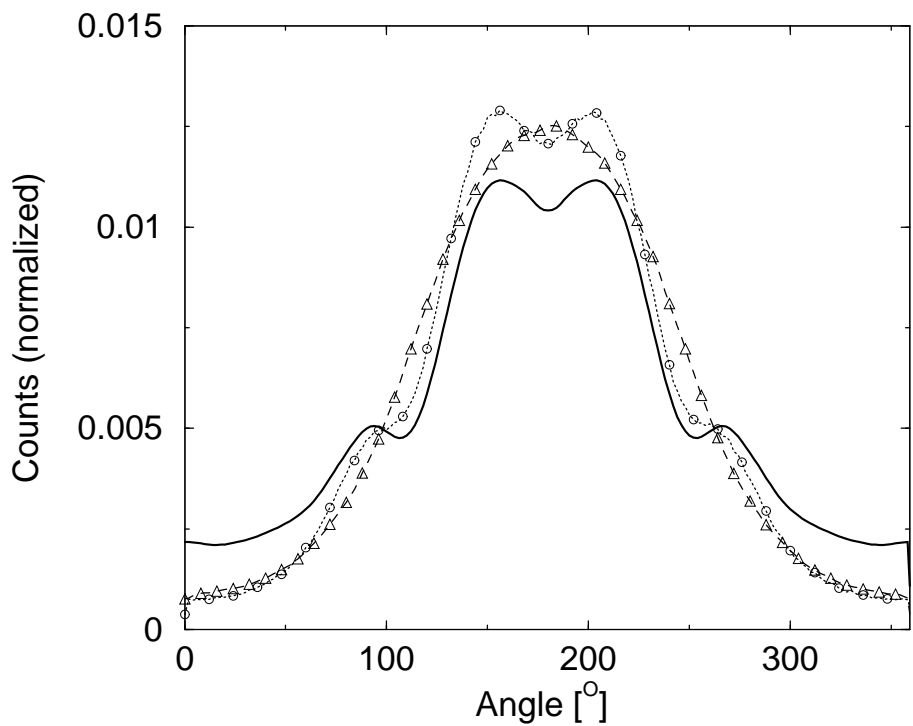}
\end{center}
\caption{Normalized distribution of the dihedral angles 
  for PAA, obtained from the simulation. The atomistic data (solid line) 
  is symmetrized and 7-point 
  smoothed. The coarse grained data is obtained for the force fields without 
  (circles) and with (triangles) torsional potential. Qualitatively, they
  reproduce the atomistic curve but for both, the peak is overrepresented 
  whilst the valley is underrepresented.}
\label{torsions}
\end{figure}

\begin{figure}[hb]
\begin{center}
\epsfxsize 12cm \epsfbox {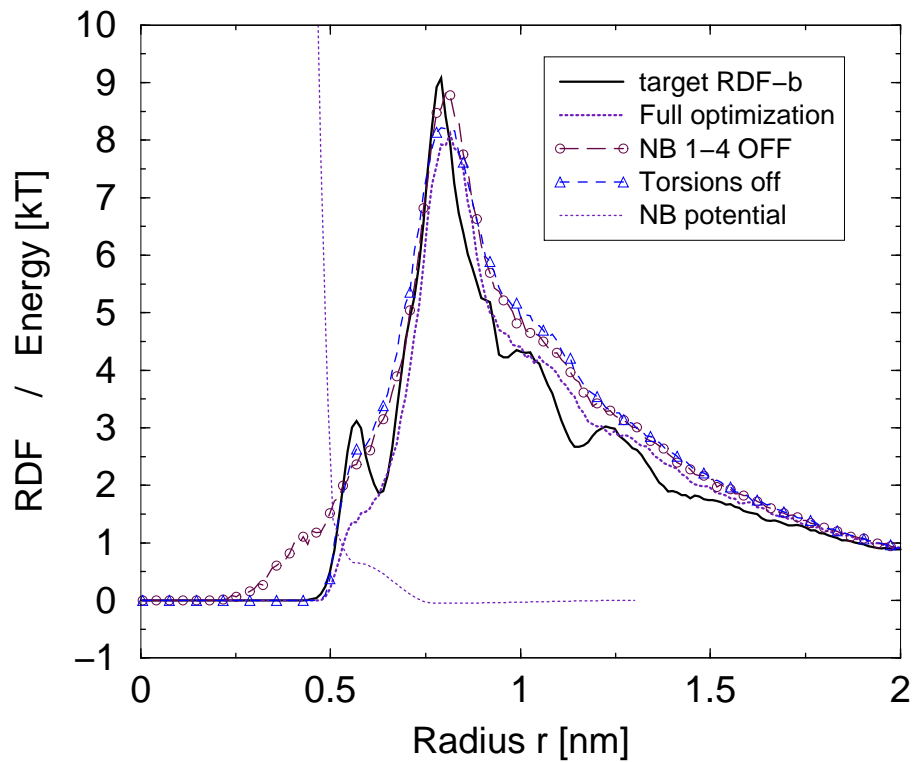}
\end{center}
\caption{Interplay of torsional and non-bonded interactions for an RDF-b of
  PAA. Runs with partly switched off interactions can be compared to the 
  full optimization. Without a non-bonded potential, the ascent starts at too 
  small disances whereas without torsions, the descent is not well reproduced.}
\label{tors_comp}
\end{figure}

A first test is the torsional output distribution of the CG simulations, 
i.e.\ the first intramolecular degree of freedom not already accounted for
by a force field term.
The final force field generates a qualitative agreement for the 
torsions, as shown in Fig.~\ref{torsions}. The solid line represents the
atomistic output distribution, the line with the circles the CG distribution.
Still, the details of the original distributions can not be reproduced.
Also the difference between the peak and the valley is too large. More
importantly, we encounter the following problem when we tried to apply the
CG force field to long polymer chains: for our test chain length of 460
repeating units, we observe a collapse of the chain. It originates
from the deep attractive part of $V_{vdw}$. Interestingly, that happens 
although the radius of gyration coincides well with the atomistic one for 
chain length 23 (cf.\ Table~\ref{rad-gyr}) and although artificial 
23-subchain RDFs of the 460-mer (which corresponds to a PAA sample with 
$43250$ g/mol) almost coincide with the original CG one
(data not shown). Due to this observations, we decide to introduce torsions
to the CG force field in order to redistribute some degrees of freedom in a
way that the attractive part of $V_{vdw}$ can be lowered. 
Therefore, the following corrections are applied: first, the depth of the
attractive tail characterized by $\varepsilon_{4}$ is lowered by one 
order of magnitude and, second, the atomistic torsional output distribution 
(cf.\ Fig.~\ref{torsions}) is used
as additional input potential in the CG picture. The optimization has to be 
corrected accordingly. Tests show at once, that torsions have a strong 
influence on the RDF-b. Figure \ref{tors_comp} shows an RDF with a fully
optimized potential compared to runs with some interaction switched off.
Without a non-bonded potential, the ascent of the main peak starts at too 
small distances. That means that the torsional potential is not strong 
enough to straighten the chain sufficiently. Still, the pre-peak is better
reproduced than in the full optimization case. If, conversely, the torsional 
potential
is taken away, the pre-peak is even better reproduced. This shows, that also
the torsions help to straighten the chain. Additionally, they make the main
peak more pronounced which can be deduced from the broadened descent without
them. Most importantly, they are capable of correcting the long-chain 
structure. Re-runs of the 460-mer prove that
the CG PAA behaves like a chain in a good solvent. We find a hydrodynamic 
radius of $R_{H} = 5.31\pm0.26$ nm for this system, which compares rather 
well with an experimental value of $R_{H} = 5.27\pm0.60$ nm for a PAA sample 
of $36900$ g/mol (corresponding to a 393-mer). 

\section{Discussion and Conclusions}
This work demonstrates the extension of the methods developed in Ref.\
\cite{meyer00} to polymers where bonded as well as non-bonded
interaction parameters have to be determined.
The simplex optimization method works also for oligomers under very 
different conditions with respect to density and charge state.
In the dense melt one has to increase the equilibration times
with respect to the simple liquids of Ref.\ \cite{meyer00}.
For poly (vinyl alcohol), it was also shown how to optimize structural and 
thermodynamic quantities at the same time. Including the pressure in the
merit function has led to a coarse grained potential with attractive as 
well as repulsive interactions. Therefore, the force field can be used in
constant-pressure simulations or to model e.g.\ surfaces.
The Boltzmann-inverted angular 
distribution was, in combination with the development of a specific 
non-bonded potential, sufficient the reproduce the target function(s). 
For poly (acrylic acid), the situation is more complicated. Here, several 
important conclusions can be drawn. First
of all, we proved that the simplex algorithm is also capable of optimizing
bonded potential variables. For this purpose, partial RDFs were utilized. 
Their peaks could be assigned to interactions between specific intramolecular 
monomers which were separately optimized. 
The torsional potential was especially important to generate the correct 
long-chain behaviour. That is, because
it allowed a successful non-bonded optimization with a fairly weak
attractive part compared to the force field derived without it. Comparison
with experimental data delivered very promising results for the hydrodynamic
radius as a first test property. 

This work will be continued. The simplex optimization is a very effective
tool for generating reliable coarse-grained force fields if one adds
human intuition to create meaningful start guesses. Special non-bonded 
spherical potentials turned out to be very useful to resolve double peaks,
without having to resort to complicated ellipsoidal potentials.
The speed-up of the simulated time compared to atomistic simulations is
enormous. This will make it possible to study properties of polymers which
are entropically dominated and which are far too long-ranged to be treated
by brute-force atomistic simulation. At the same time, the parameterization
procedure ensures that enough of the chemical identity of the polymer is
kept to allow investigations and deeper understanding of specific systems.
The PVA force field is intended to be used for equilibrating large samples 
with surfaces.
The PAA force field shall be utilized to simulate several long PAA strands.
That will indicate the transferability of the model and a thorough test 
against experiments. 

After completing this paper we learned about some interesting related work
\cite{akkermans}. The authors proceed according to the same idea of
optimizing the interactions of coarse-grained polymer models with
respect to structural properties.  However, their method and aim are
quite different: Using a Monte Carlo sampling method they optimize
interaction parameters on a much coarser level of blobs representing
larger subchains. While this might be useful for melt simulations,
their procedure abandons too much details for our interest in
surface properties.

\section*{Acknowledgements}
We would like to thank Beate M\"uller and Simone Wiegand for experimental
PAA measurements of the hydrodynamic radius.
Oliver Biermann is acknowledged for making available his atomistic PAA data
as well as for technical improvements of the simplex optimization and 
Oliver Hahn for contributing an MD program for chains of spherical particles.
%
%


\clearpage

\begin{thebibliography}{10}

\bibitem{baschnagel00}
J.~Baschnagel, K.~Binder, P.~Doruker, A.~A.~Gusev, 
O.~Hahn, K.~Kremer, W.~L.~Mattice,
F.~M{\"u}ller-Plathe, M.~Murat, W.~Paul,
S.~Santos, U.~W.~Suter and V.~Tries, 
\newblock Adv. Polym. Sci. {\bf 152}, 41 (2000).

\bibitem{forrest95}
B.~Forrest and U.~W.~Suter,
\newblock J. Chem. Phys. {\bf 102}, 7256 (1995).

\bibitem{carmesin88}
I.~Carmesin and K.~Kremer,
\newblock Macromolecules {\bf 21}, 2819 (1988).

\bibitem{paul91}
W.~Paul, K.~Binder, K.~Kremer, and D.~Heermann,
\newblock Macromolecules {\bf 24}, 6332 (1991).

\bibitem{baschnagel91}
J.~Baschnagel, K.~Binder, W.~Paul, M.~Laso,
U.~W.~Suter, I.~Batoulis, W.~Jilge
and T.~ B{\"u}rger, 
\newblock J. Chem. Phys. {\bf 95}, 6014 (1991).

\bibitem{doruker97}
P.~Doruker and W.~Mattice,
\newblock Macromolecules {\bf 30}, 5520 (1997).

\bibitem{doruker99a}
P.~Doruker and W.~Mattice,
\newblock J. Phys. Chem. B. {\bf 103}, 178 (1999).

\bibitem{groot97}
R.~Groot and P.~Warren,
\newblock J. Chem. Phys. {\bf 107}, 4423 (1997).

\bibitem{espanol97}
P.~Espanol, M.~Serrano, and I.~Zuniga,
\newblock J. Mod. Phys. C {\bf 8}, 899 (1997).

\bibitem{faller99}
R.~Faller, H.~Schmitz, O.~Biermann, and F.~M\"uller-Plathe,
\newblock J. Comput. Chem. {\bf 20}, 1009 (1999).

\bibitem{meyer00}
H.~Meyer, O.~Biermann, R.~Faller, D.~Reith, and F.~M\"uller-Plathe,
\newblock J. Chem. Phys. accepted (2000).

\bibitem{tschoep98a}
W.~Tsch{\"o}p, K.~Kremer, J.~Batoulis, T.~B{\"u}rger, and O.~Hahn,
\newblock Acta Polymer {\bf 49}, 61 (1998).

\bibitem{tschoep98b}
W.~Tsch{\"o}p, K.~Kremer, O.~Hahn, J.~Batoulis, and T.~B{\"u}rger,
\newblock Acta Polymer {\bf 49}, 75 (1998).

\bibitem{eilhard99}
J.~Eilhard, A.~Zirkel, W.~Tsch{\"o}p, O.~Hahn,
K.~Kremer, O.~Scharpf, D.~Richter
and U.~Buchenau, 
\newblock J. Chem. Phys. {\bf 110}, 1819 (1999).

\bibitem{All87}
M.~Allen and D.~Tildesley,
\newblock {\em {Computer Simulation of Liquids}},
\newblock Oxford Science, Oxford, 1987.

\bibitem{grest86}
G.~Grest and K.~Kremer,
\newblock Phys. Rev. A {\bf 33}, 3628 (1986).

\bibitem{MPvG97poly}
F.~M\"uller-Plathe and W.~F. van Gunsteren,
\newblock Polymer {\bf 38}, 2259 (1997).

\bibitem{yasp}
F.~M{\"u}ller-Plathe,
\newblock Comp. Phys. Comm. {\bf 78}, 77 (1993).

\bibitem{MP98jms}
F.~M\"uller-Plathe,
\newblock J. Membrane Sci. {\bf 141}, 147 (1998).

\bibitem{biermann00}
O.~Biermann and F.~M\"uller-Plathe,
\newblock in preparation  (2000).

\bibitem{akkermans}
R.~L.~C. Akkermans,
\newblock {\em A structure-based coarse-grained model for polymer melts},
\newblock PhD thesis, University of Twente (Netherlands), 2000,
\newblock chapter 5.

\end{thebibliography}

%
%
%
%
%
%
\clearpage

%
%
\begin{table}[htbp]
  \begin{center}
    \caption{Fit results (of two Gaussians to the data) for the atomistic 
      bending angle distribution as well as optimized parameters for the CG 
      model of the PAA center-of-mass repeat unit [possibility 2].}
    \begin{tabular}{cccc}
       &            &          & \\
Peak \# & position & standard deviation & peak height\\
       & [$^{o}$]   &  [$^{o}$] & [height of peak 2]\\
       &            &          &  \\ \hline\hline
Atomistic fit results &&&\\
1 & 88.5 & 10.7 & 1.72\\
2 & 118.2 & 11.6 & 1.00\\[10pt] \hline
force field without torsional potential&&&\\
1 & 88.5 & 6.8 & 1.90\\
2 & 116.7 & 7.4 & 1.00\\[10pt] \hline
force field including torsional potential&&&\\
1 & 88.0 & 6.8 & 1.72\\
2 & 116.7 & 7.4 & 1.00\\[10pt]
    \end{tabular}
    \label{table-bond-parms}
  \end{center}
\end{table}

\newpage

\begin{table}[htbp]
  \begin{center}
    \caption{Optimized non-bonded parameters for the coarse grained PAA
      simulations.}
    \begin{tabular}{ccccccccc}
& $\sigma_{1}$ & $\varepsilon_{1}$ & $\sigma_{2}$ & $\varepsilon_{2}$ %
& $\sigma_{3}$ & $\varepsilon_{3}$ & $\sigma_{4}=r_{cut}$ & $\varepsilon_{4}$ %
\\[-10pt] 
& & & & & & & & \\ \hline\hline
force field without torsional potential& & & & & & & \\
& 0.507 & 11.3 & 0.569 & 0.390 & 0.790 & 2.31 & 1.3 & 0.50\\[10pt] \hline
force field including torsional potential& & & & & & &\\
& 0.496 & 11.3 & 0.559 & 0.353 & 0.775 & 0.49 & 1.3 & 0.05\\[10pt]
    \end{tabular}
    \label{opt-nb-parms}
  \end{center}
\end{table}

\newpage

\begin{table}[htbp]
  \begin{center}
    \caption{Radius of Gyration [nm] for PAA 23-mers. 
      Comparison between atomistic and coarse grained chain.}
    \begin{tabular}{ccc}
        $R_{G}^{atomistic}$ & $R_{G}^{CG}$ without torsions & $R_{G}^{CG}$ %
        including torsions\\
        &&  \\ \hline\hline
        $1.28\pm0.08$ & $1.24\pm0.05$ & $1.25\pm0.05$ \\
    \end{tabular}
    \label{rad-gyr}
  \end{center}
\end{table}

%
%
%
%
%
%
\clearpage

\listoffigures

\newpage

\newpage

\newpage

\newpage

%
%
\newpage

\newpage

\newpage

%
%
\newpage

\newpage

\newpage

\end{document}